\documentclass[runningheads]{llncs}
\usepackage{graphicx}
\usepackage[table]{xcolor}
\begin{document}
\title{Digital Twin of a Cloud Data Centre: OpenStack Cluster Visualisation
}
%
%
\author{Sheridan Gomes\and
Adel N. Toosi\orcidID{0000-0001-5655-5337}\and
Barrett Ens\orcidID{0000-0001-6695-4809}}
%

%
\institute{Faculty of Information Technology\\ Monash University, Clayton, Australia\\
\email{\{adel.n.toosi,barrett.ens\}@monash.edu}\\}
\maketitle              
\begin{abstract}
Data centres in contemporary times are essential as the supply of data increases. Data centres are areas where computing systems are concentrated for facilitating data processing, transfer and storage. At present traditional data centres have moved more towards the cloud model—thereby making the processing, storage and harnessing of data more manageable and more accessible via the utility and subscription-based model of computing services. From the administrative point of view, cloud data centres are complex systems, hard to grasp and require large amounts of time to analyse different aspects of the cloud data centre such as maintenance and resource management. For a cloud data centre admin, this could be a challenging problem and a highly time-consuming task. Accordingly, there is a need to improve the useability of cloud data centre monitoring and management tools, and the digital twin could fulfil this need. This paper's primary objective is to construct a digital twin — a 3D visualisation and monitoring tool — of a cloud data centre managed by OpenStack, the well-known open-source cloud computing infrastructure software. To evaluate our proposed tool, we garner feedback on the digital twin's useability compared to the OpenStack dashboard. The input will be received from cloud data centres experts as they test the digital twin and answer various questions in an interview. The study results show that our proposed Digital Twin will help data centre admins better monitor and manage their data centres. It also will facilitate further research and implementation of the digital twin of data centres to improve usability.

\end{abstract}
\section{Introduction}
The beginning of the modern era has seen an exponential rise in Information Technology (IT). The early 21st century saw the rise and growth of mobile phones, computers and tablet computers. These devices have now become a staple in households, academia and business enterprises all around the world. As we progress through this age, technology has advanced further and further, giving rise to new Internet-of-Things (IoT) devices ranging from smart fridges and smart TVs, to even smart weight scales, to name a few. All these devices, along with IoT and other IT applications, produce massive amount of data. Data in today's world is one of the most critical and essential resources for companies and people alike. Data centres are built to meet this demand for data storage and analysis, especially cloud data centres where data is stored, accessed and analysed on the cloud. Cloud data centres are computational infrastructures available globally and provide services that allow the storage, access, and analysis of data on remote servers. Cloud data centres have numerous advantages compared to their conventional counterpart and provide a highly scalable and cost-effective way of accessing computational resources, often in the form of virtual machines (VMs). OpenStack is an open-source cloud computing platform, mostly deployed as infrastructure-as-a-service in both public and private clouds where VMs and other resources are made available to users~\cite{fifield2014openstack}. 

While existing management and monitoring tools such as OpenStack dashboard (Horizon) are precious for data centre administrators, data centre management and monitoring present several challenges for the operators~\cite{kumar2014open}. Data centre admins need to have real-time and unified information about their data centre's resources and health status to make quick decisions. For example, in the case of server maintenance where repairs are required on a physical server, it is essential to quickly find a suitable host to accommodate tenants' VMs running on the failing server. In addition, data tables or text table view of data provided by tools such as Horizon often fail to deliver admins' all essential requirements. For example, more detailed data about the exact location of a physical server or its power consumption might be required and tools like OpenStack fail to provide that in their dashboard.
Finally, working with the OpenStack dashboard requires a high amount of knowledge and expertise to understand and analyse the dashboard's different aspects. Further, because of how the human brain processes information, visualisation tools can represent large amounts of complex data such as data centre resources much more straightforward than tables or raw data. This research paper aims to represent a visualisation tool to help data centre admins better perform their monitoring and management tasks.
 
In this paper, we propose a digital twin of the cloud data centre to tackle the above complexities. A digital twin is a virtual representation of an object. A digital twin can represent different aspects of a system in an easy to understand virtual setting, which allows for higher usability and a better understanding of the whole system. Digital twins have been used to handle many complex systems ranging from aviation to manufacturing~\cite{ayani2018digital}. A user-friendly and time-saving monitoring tool for cloud data centres would positively impact the industry by reducing complexity and increasing productivity in resource monitoring and management, which would be essential for a cloud data centre administrator. For testing and research on the proposed hypothesis, a digital twin prototype was developed using Unity3D software. 
Unity3D was used to develop the prototype as it is easy to use, works well with API's and has a great support community. 
The digital twin needed to be implemented to work with OpenStack APIs and some other infrastructure level monitoring APIs to synchronise the digital twin with the cloud data centre in real-time. Any changes made to the cloud data centre would be reflected in the virtual environment on the digital twin instance and vice versa. 

Once the digital twin prototype is developed using Unity 3D to ascertain its usability as a monitoring and management tool, 
an interview was conducted with multiple participants who are experts in cloud data centres. These participants had to watch a live demonstration of the digital twin in action, illustrating the digital twin's different features. An interview was conducted asking questions to gauge how the digital twin would improve the data centre admin experience compared to the OpenStack dashboard if various fields such as usability, complexity and time management are considered.  The feedback from the experts helped us understand the benefits, differences and future need for improvement. 

The \textbf{main contributions} of the paper are as follows:
\begin{itemize}
\item A new brief taxonomy of digital twin characteristics and elements.
\item The design and implementation of a digital twin — a 3D visualisation, monitoring, and management tool — of a cloud data centre using Unity3D. To the best of our knowledge, the proposed digital twin of a cloud data centre is one of its first kind in the literature. The digital twin uses data fed from the OpenStack API's and infrastructure level devices such as Enclosure Power Distribution Units (ePDUs) to provide an interactive and real-time virtual representation of the different aspects such as VMs, hypervisors and physical servers of the cloud data centre running on OpenStack platform.
\item We provided insights for future directions for building a more advanced digital twin of a data centre based on the expertise and knowledge gained from conducting the research and implementation of our cloud data centre digital twin.
\end{itemize}
The rest of the paper is organised as follows: Section~\ref{sec:back} outlines the related works, background and taxonomy of digital twin characteristics. In Section~\ref{sec:model}, the research methodology is explained, this includes the system design of the digital twin, implementation, prototyping and interview process. Section~\ref{sec:study} describes the findings and discussions of the research and interviews. Lastly, we conclude with the summary of the research conducted and highlight some future works.

\section{Background} \label{sec:back}
Digital Twin technology is described as a virtual representation of a tangible real-world object. The digital twin has two primary components: the physical twin (the real-world physical object) and the virtual twin (a virtual representation of the physical object). The primarily related works of the digital twin are in the aviation and manufacturing industries, with potentials in other sectors being high. A digital twin of a cloud data centre is a new idea, and there is not much research related to that topic~\cite{boschert2016digital}. To the best of our knowledge, we are the first to propose a non-commercial digital twin of a cloud data centre with features discussed here.   

\subsection{A taxonomy of digital twin characteristics}
\begin{figure*}[htpb!]
\centering
\includegraphics[width=0.7\textwidth]{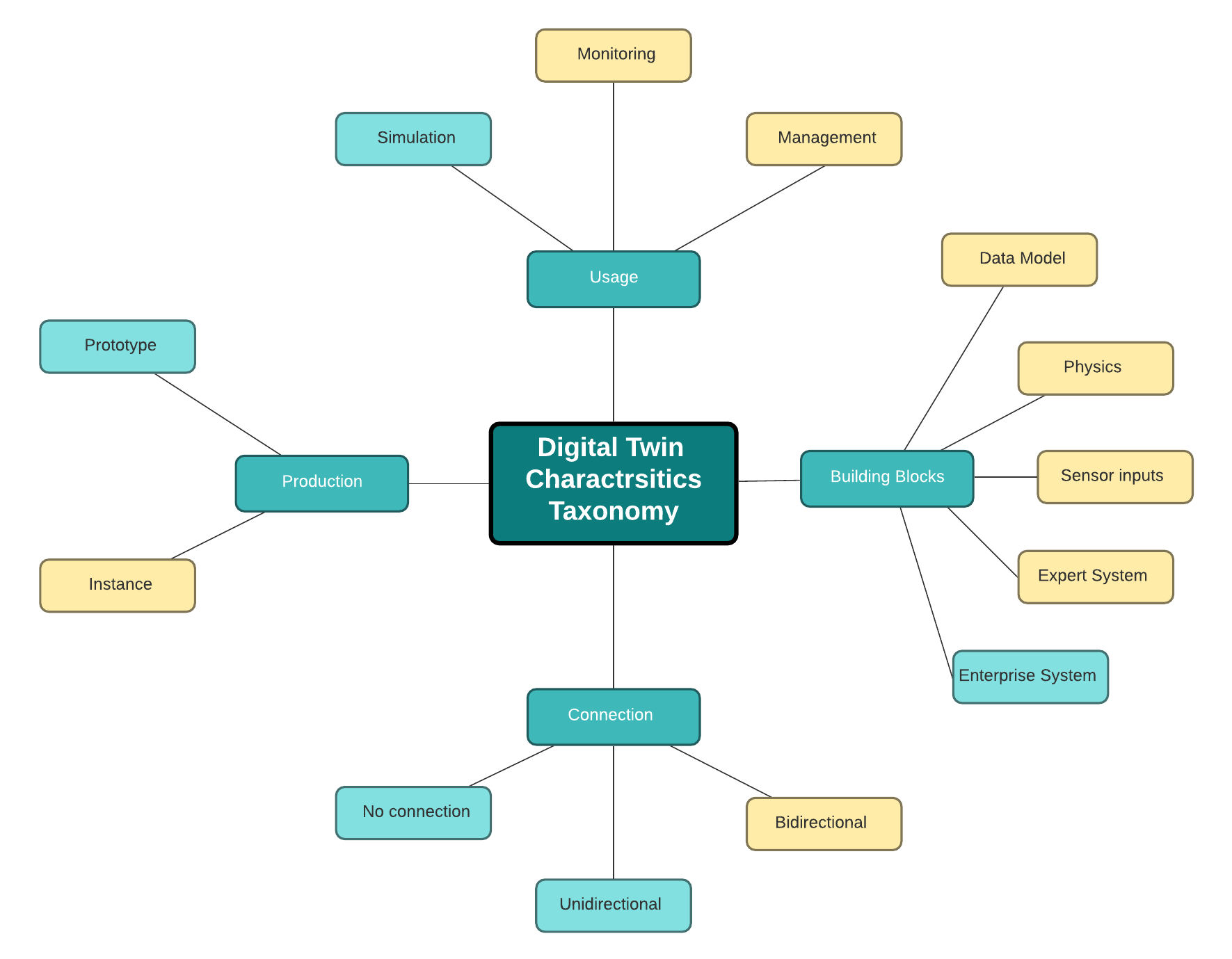}
\caption{Taxonomy of Digital Twin Characteristics}
\label{fig:taxo}
\vspace{-1.2cm}
\end{figure*}

Figure~\ref{fig:taxo} provides a taxonomy of digital twin characteristics. In the following section, we explain each element of the taxonomy and map our model to that.

\subsubsection{Digital twin usage.} The usage of the digital twin can be classified into three kinds of project dimensions: \textit{simulating}, \textit{monitoring} and \textit{managing}. Replication/simulation digital twin dimension is where the digital twin projects main goal is to reproduce the physical object and simulate various aspects to ensure that the physical object is ready to be deployed without any issues. The monitoring dimension of the digital twin is when the digital twin is used to monitor a physical object; this can be done to ensure that the physical item is working correctly and as intended~\cite{8632888}. Lastly, the digital twin's managing dimension is when the digital twin's primary aim is to manage the physical item, which would mean that the physical twin and the digital twin would have a bi-directional connection or two-sided connection. Wherein the data flows from the physical twin to the digital twin and vice versa, allowing the digital twin to manage the physical twin~\cite{inproceedingsDroder}. Instances of digital twin utilisation are a digital twin represented by Mohammadi and Taylor~\cite{8285439}, who created a digital twin of a city to visualise the different aspects of the town such as roads; this would help in town planning and management. Another instance of a digital twin used in urban planning was the digital twin represented by Lieberman et al.~\cite{lieberman2017using}. They developed a digital twin used to create an underground city plan. Please note that the digital twin of a cloud data centre we propose in this paper is designed to be used for monitoring and managing purposes. 

\subsubsection{Period of production.}  The period of production for the digital twin can be classified into two parts. Digital twins developed before the creation of the physical twin, which are called \textit{digital twin prototypes}, and digital twins created along with or after the creation of the physical twin called \textit{digital twin instances}~\cite{grieves2015digital}. Ayani et al.~\cite{ayani2018digital} represented an example of a digital twin prototype; they used it throughout the configuration phase, which allowed their clients to give feedback and helped avoid costs related to prototyping. An example of a digital twin instance was represented by Luo et al.~\cite{luo2018digital}. They created a digital twin of a milling machine to predict wear and tear and other issues to the milling machine under different conditions. In our proposed model, we develop a digital twin instance of a cloud data centre.

\subsubsection{Connection between physical twin and digital twin.} Another characteristic of the digital twin is its connection with its physical counterpart. There are three ways of classifying the connection between a digital twin and a physical twin representing in what way the two parts are linked. Firstly, \textit{no connection}, wherein the digital twin and physical twin are not connected and have no association with one another. Secondly, a \textit{unidirectional connection}, wherein the digital twin and the physical twin have a one-sided relationship where data from the physical twin is fed to the digital~\cite{inproceedingsDemkovich}. An example of this is represented by Luo et al. ~\cite{luo2018digital}, where they use a unidirectional relationship to predict issues with the device used. Lastly, a \textit{bidirectional connection} between a digital twin and a physical twin, where the data from the digital is fed to physical and vice versa, thereby creating a dual-link that data is fed back and forth. For instance, this is represented by Malik and Bilberg~\cite{malik2018digital}, where they use a bidirectional connection to control a robot using its digital twin counterpart. Please note that our proposed digital twin of a cloud data centre is designed to have a bidirectional connection with the physical twin. 

\subsubsection{Building blocks of a digital twin.} Digital twins can vary in complexity depending upon features and input data. The classification is divided into five levels, with each level increasing in complexity and input data sources. Level one only has one input data source, which is the data model~\cite{tao2017digital}. Level two digital twin where physics are added to the data model of the level one digital twin creating a more elaborate digital twin. The third level, where a level one digital twin is incorporated with sensor data to increase the productivity of the model~\cite{9120192}. Level four has the level three digital twin enhanced with physics and expert systems. Level five, where enterprise systems are added to a level four digital twin~\cite{articleKritzinger}.

%

\section{Digital twin of a cloud data centre instance system}\label{sec:model}
\begin{figure}[htpb!]
\centering
\includegraphics[width=0.8\columnwidth]{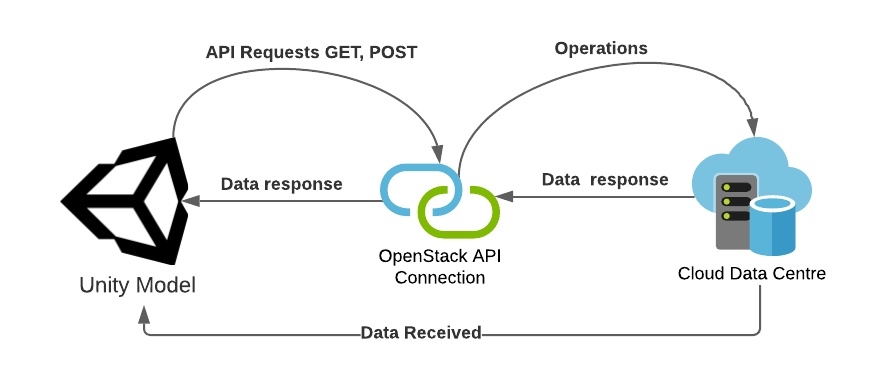}
\vspace{-0.5cm}
\caption{System Overview}
\label{fig:second}
\vspace{-1cm}
\end{figure}

To develop the digital twin instance, decisions needed to be made on various design options and other system choices on how the system would look and be developed. The proposed system consists of two main parts the \textit{unity model} and \textit{APIs integration} (Figure~\ref{fig:second}). The two parts of the system need to work in unison to represent the cloud data centre effectively. 

\subsection{Unity Model} 
One of the primary components of the digital twin is the unity model, which would help visualise the cloud data centre in Unity 3D software. The Unity 3D software is used for developing full-fledged games for both 2D and 3D. It offers various resources and an abundance of assets primarily used for 3D modelling~\cite{xie2012research}.

\subsubsection{Main Components} The unity model's design is an integral part of the digital twin system. To ensure that the digital twin represents all the essential elements of the cloud data centre, we analysed the different parts of the cloud data centre and put together a model to ensure that the design standards are met. The main components of the cloud data centre that need to be represented in the digital twin are:

\begin{itemize}
\item \textit{Hypervisors (physical servers)}: OpenStack supports software that allows the user (data centre admin) to manage physical servers by accessing the underlying hardware component. Hypervisor software gives the user the ability to manage and control the VMs hosted on a particular server~\cite{rosado2014overview}. In our proposed unity model, these hypervisors (physical servers) are represented as plates (Figure~\ref{fig:placement}). We leveraged a small-scale OpenStack managed cluster at DisNet Laboratory in Monash University to build the digital twin. These servers are classified as hypervisors on the system and are real-world tangible objects. The size of the plate is proportional to the number of VCPUs (virtual CPUs) it supports. Since hypervisors in compute nodes support 32 VCPUs, each plate's dimensions are x = 8 and z = 4, where x is the width and z the length. As it is depicted in~Figure~\ref{fig:plate}, each plate is divided into 32 grid sections, each representing a VCPU. Energy consumption of physical servers are also shown based on the darkness of colours from light red to dark red to denote energy consumption from low to high, respectively. We use existing ePDUs in the cluster to read live metering data including power consumption in each server.

  




\begin{figure}
\begin{minipage}[c]{0.6\linewidth}
\includegraphics[width=1\columnwidth]{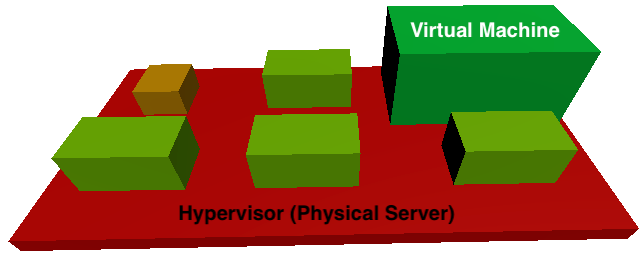}
\caption{Box (Virtual Machine) Placement on the plate (Hypervisor)}
\label{fig:placement}
\end{minipage}%
\vspace{-0.5cm}
\hfill
\begin{minipage}[c]{0.4\linewidth}
\centering
\includegraphics[width=1\columnwidth]{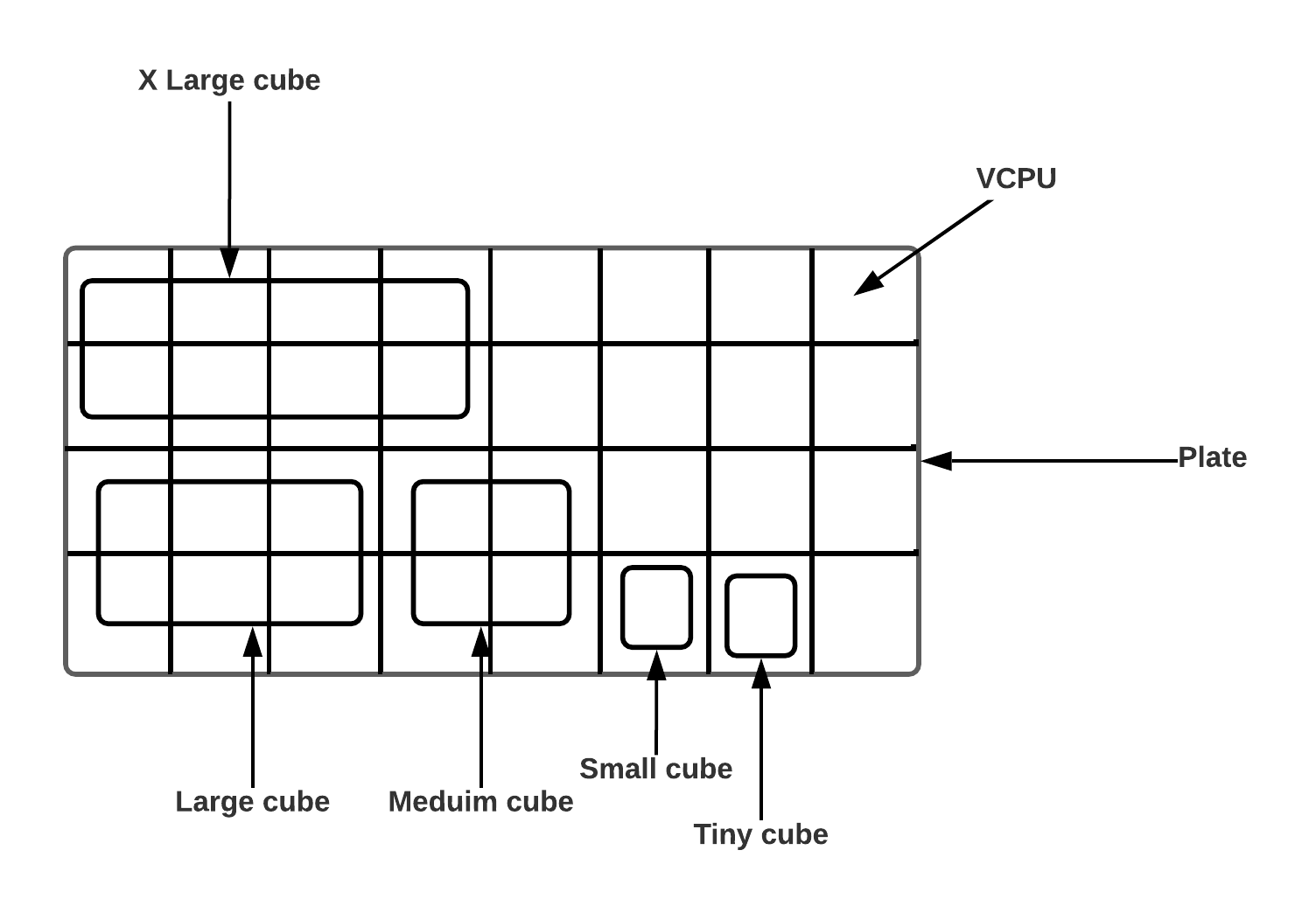}
\caption{Plate configuration}
\label{fig:plate}
\end{minipage}
\end{figure}


\item \textit{Virtual Machines} (VMs): VMs are emulating a computer system that is created, stopped or terminated, and hosts an operating system (OS) to execute user's applications. In OpenStack, VMs are called \textit{instances}~\cite{khan2011openid} and in the OpenStack APIs they are called \textit{servers}. In our implementation, we represented VMs as boxes similar to the example shown in Figure~\ref{fig:placement}. To be a complete digital twin of the cloud data centre, the unity model has the boxes (VMs) placed upon the plates (hypervisors), giving a virtual, abstract and simplistic view of how VMs run on a real-world server. 
The placement of the boxes on the plate is essential for proper visualisation of the cloud data centre. 
Our proposed placement algorithm places places the boxes on the plate until the sum of the boxes' size has reached the limit of the x-axis (right to left of the plate), in our case that is 8. The algorithm moves to the next row and changes the value of the z-axis for each box, the process then repeats itself until all VMs are placed. Figure~\ref{fig:placement} depicts a sample placement.



Similar to plates, we use transparency to show a VM is suspended or shut off. The system displays boxes in a semi-transparent way to reflect that a VM is suspended or shut off in the digital twin. Figure~\ref{fig:powerstate} illustrates power on and off VMs as semi-transparent and solid boxes, respectively.

\vspace{-0.5cm}
\begin{figure}[htpb!]
\centering
\includegraphics[width=0.4\columnwidth]{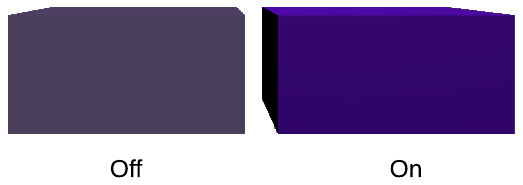}
\caption{Power State of a VM shown through transparency.}
\label{fig:powerstate}
\vspace{-0.5cm}
\end{figure}

\item \textit{Flavours}: The user can configure various aspects of the instance, such as the number of VCPUs, memory and storage~\cite{rosado2014overview}. These aspects are represented on the digital twin as the boxes' size. Figure~\ref{fig:flavours} shows different flavours represented by various box sizes. The size of the base area of the box is proportional to the number of VCPUs (see Figure~\ref{fig:plate}), and the volume of the box is proportional to the size of the main memory. For simplicity and better visualisation, we do not represent the disk size of the VM in the model. OpenStack, by default, supports five different flavours, as shown in Table~\ref{tab:flavors}. 


\vspace{-0.5cm}
\begin{figure}[htpb!]
\centering
\includegraphics[width=0.8\columnwidth]{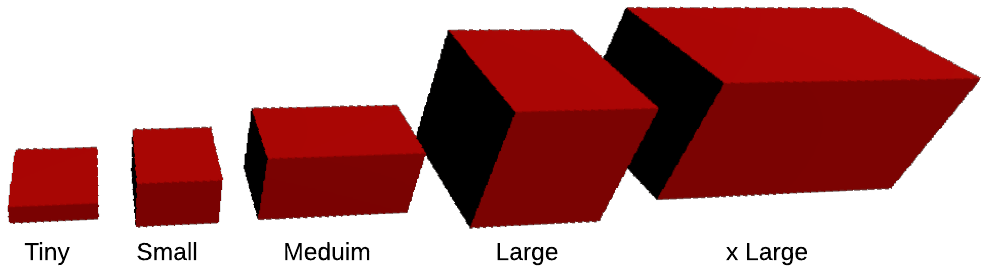}
\caption{Flavours are represented by various box sizes}
\label{fig:flavours}
\vspace{-0.5cm}
\end{figure}


\vspace{-0.5cm}
\begin{table}[hptb!]
\scriptsize
\centering
\caption{\label{tab:flavors} OpenStack Default Flavors Details.}
\begin{tabular}{|l|l|l|l|}
\hline
Flavor    & VCPUs & Disk (in GB) & RAM (in MB) \\ \hline
m1.tiny   & 1     & 1            & 512         \\ \hline
m1.small  & 1     & 20           & 2048        \\ \hline
m1.medium & 2     & 40           & 4096        \\ \hline
m1.large  & 4     & 80           & 8192        \\ \hline
m1.xlarge & 8     & 160          & 16384       \\ \hline
\end{tabular}%
\vspace{-0.5cm}
\end{table}
  
\item \textit{Projects}: Each instance belongs to a user project or tenant. In the proposed digital twin, the instances that belong to a certain projects are all represented in the same color and with different colours for each project. 

\end{itemize}

\subsubsection{Digital twin interactive features}
The digital twin also offers a couple of \textit{interactive features} to demonstrate the possibility of cloud data centre management and having the ability to perform interactive operations.

\textit{Start and shut off virtual machines and physical servers:}
The user can shut off or power on a VM or a physical server by long right-clicking on a box or plate, respectively. The box/server will start blinking during the time the switching up or down operations is happening. To know if the VM has successfully turned on or off the unity model checks with OpenStack every second and compares the old status of the VM to the new status when the status changes the whole model is instantly refreshed and shows the new changes. Similar process will be done for the physical servers (hypervisors). 

\textit{Virtual machines migration:} 
To migrate a VM from one hypervisor to another, the user might drag and drop a box from a plate to another. When the migration is happening, the box keeps blinking until the migration is completed, once migration is over the box stops blinking, and the model is updated.

\begin{figure}[htpb!]
\centering
\includegraphics[width=0.4\columnwidth]{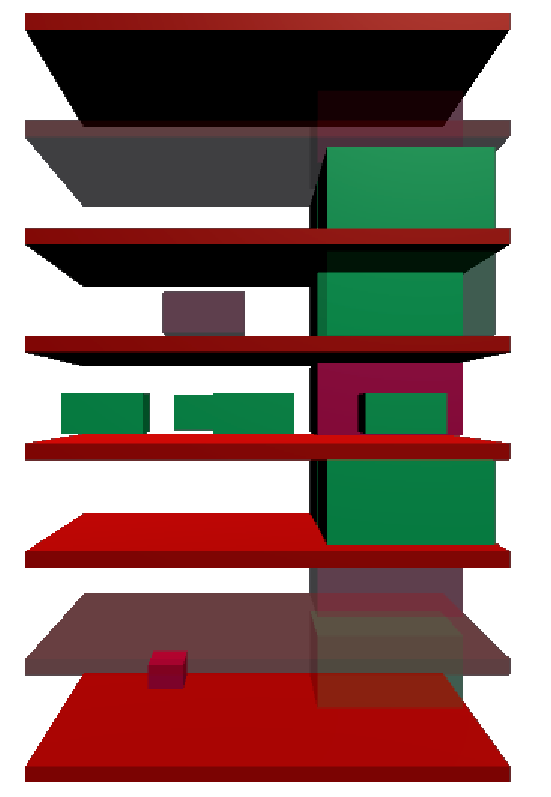}
\caption{Complete model}
\label{fig:sixth}
\vspace{-0.5cm}
\end{figure}

Figure~\ref{fig:sixth} illustrated a complete unity model and displays how the digital twin represents a rack of servers in a cloud data centre in a 3D visualisation model.

\subsection{APIs integration}
OpenStack provides a full range of different RESTful GET APIs to receive data from the cloud data centre and POST APIs which would allow the Unity model to interact with the cloud data centre thereby ensuring that the digital twin would have a bidirectional relationship to its counterpart~\cite{khan2011openid}. 

\begin{itemize}
\item{GET Requests}: The digital twin instance uses GET requests to fetch data from the cloud data centre and for monitoring purposes. The required data includes a list of servers (VMs or instances), list of hypervisors, list of flavours and list of projects. 


\item{Post APIs}: The Post requests are needed to implement management and operational features of the digital twin. A POST request is used to manipulate data on the server-side or pass data from the digital twin to the server using API endpoints. For the interactive elements of the digital twin, we need to call a couple of POST requests. To perform authentication and authorisation, a POST request is sent to OpenStack's keystone service, which returns an authorisation token. This token needs to be used in all request headers for future GET and POST requests.
\end{itemize}


\section{Qualitative Study}\label{sec:study}
As part of the research process, an interview was held. The interview session's main aim was to get feedback and evaluate the effectiveness of the digital twin system compared to other existing monitoring and management tools, specifically the OpenStack dashboard. OpenStack dashboard was chosen for the comparison as it is the primary tool used to monitor servers in our cluster.  

\subsubsection{Study Procedure.} The study consists of a semi-structured interview process; this interview was conducted on participants selected on specific criteria listed below.

\subsubsection{Participants.} The participants selected for the interview process are experts in the field of cloud data centres and worked with OpenStack. The number of participants in the interview is three. The participants were mixed gender, having two males and one female participant and are selected according to years of experience, level of study and type of knowledge related to data centres and OpenStack. Table~\ref{tab:participents} shows more detail of each participant involved.

\vspace{-0.5cm}
\begin{table*}[hptb!]
\scriptsize
\centering
\caption{\label{tab:participents} Participant Details.}
\begin{tabular}{|l|l|l|l|l}
\cline{1-4}
Participants & \begin{tabular}[p]{@{}l@{}}Study level \\ (degree)\end{tabular} & \begin{tabular}[c]{@{}l@{}}Experience with \\  Openstack years)\end{tabular} & \begin{tabular}[c]{@{}l@{}}Experience with   cloud computing \\ or cloud data centres (years) (type)\end{tabular} &  \\ \cline{1-4}
P1           & PhD                                                             &  2 years                                                                            & 2 years and worked                                                                                                                  &  \\ \cline{1-4}
P2           & PhD                                                             & 3 years                                                                       & 6 years and research projects                                                                                     &  \\ \cline{1-4}
P3           & Bachelor                                                                & 8 years                                                                              & 15 years and worked (cloud architect) and projects                                                                                                                  &  \\ \cline{1-4}
\end{tabular}%
\vspace{-1.0cm}
\end{table*}

\subsubsection{Interview procedure.} First, a demonstration of the digital twin was shown to the participants. The demonstration was around 10 minutes and demonstrated all the system's functionalities, displaying and explaining all the visual components of the digital twin and interactive elements. After the demonstration, interview questions are asked. Table~\ref{tab:Qs} shows these questions. The first group of questions were asked to gauge the familiarity and knowledge of the participants with the technologies used in creating the digital twin like Unity 3D and cloud data centres. Then five questions in the second group of questions in Table~\ref{tab:Qs} were asked where the participants are instructed to answer on a Likert scale of 1 - 5 as follows: 1) very poor, 2) poor, 3) fair, 4) good and 5) very good. Lastly, the participants were asked four open-ended questions shown in Table~\ref{tab:Qs}.

\vspace{-0.5cm}
\begin{table}[htpb!]
\scriptsize
\centering\caption{\label{tab:Qs} Interview questions.}
\begin{tabular}{|l|l|}
\hline

\multicolumn{2}{|c|}{\cellcolor{blue!25} \textbf{Questions to gauge familiarity and knowledge of each participant}}                                                                                                                                                                                            \\ \hline
Q1              & \begin{tabular}[c]{@{}l@{}} How would you rate your familiarity and knowledge of \\cloud data centres and OpenStack?\end{tabular}  \\ \hline
Q2              & \begin{tabular}[c]{@{}l@{}} How would you rate your familiarity and knowledge of \\Unity3D?\end{tabular}                             \\ \hline

Q3              & \begin{tabular}[c]{@{}l@{}}How useful is the digital twin system compared to \\OpenStack dashboard for the system representation?\end{tabular} \\ \hline

\multicolumn{2}{|c|}{ \cellcolor{blue!25}  \textbf{Survey style questions}} \\ \hline

Q4              & How familiar is the digital twin system (with instructions)?                                                                                                                                         \\ \hline
Q5              & How user friendly is the digital twin system?                                                                                                                                                         \\ \hline
Q6              & Is the digital twin easy to understand?                                                                                                                                                               \\ \hline
Q7              & Is the digital twin likely to save time while monitoring?                                                                                                                                             \\ \hline
Q8              & How satisfied are you with the digital twin application?                                                                                                                                              \\ \hline
\multicolumn{2}{|c|}{  \cellcolor{blue!25}  \textbf{Open-ended questions}} \\ \hline
Q9              & What   changes or additional features would you recommend? \\ \hline
Q10             & What   are the negative aspects of the digital twin?       \\ \hline
Q11             & What   are the positive aspects of the digital twin?       \\ \hline
Q12             & Any Additional feedback?                                   \\ \hline
\end{tabular}%
\vspace{-1.0cm}
\end{table}


\section{Evaluation and Results} As it is shown in Table~\ref{tab:results}, the participants involved rated their familiarity and knowledge of cloud data centre and OpenStack technologies (Q1) highly, with P1, P2 and P3 giving a score of 4, 4, and 5 respectively. The scores indicate that the participating interviewees have an insight into the technology used and can provide valuable feedback regarding that. The participants were also asked about their knowledge and familiarity with Unity 3D (Q2), with P1, P2 and P3 giving a rating of 1, 3, and 2 respectively. These scores show that the participants are not very familiar with Unity 3D; this adds an element of usability where the user does not have to have experience with unity to understand or use the system.
The participants also answered five survey style questions. Their answers highlight the usability, efficiency in terms of time-saving, and familiarity with the system. For question Q3, all the participants rated the digital twin system's usability relatively high with 4, 4, and 3 for P1, P2 and P3, respectively. Next question Q4 was a question regarding the dimension of familiarity; here, two participants rated the system positively (4 and 5). However, P3 rated a score of 2 where he/she believed that improvements could be made on the side of familiarity by adding more instructions. For question Q5, all the participants gave an overwhelmingly high score, which demonstrates that the digital twin is more user friendly and can be used even by people who might not possess much knowledge of cloud data centres. Question Q6 was also rated highly by the participants, showing that the digital twin performs well in being easy to understand. The next question Q7 dwells into how well the digital twin would be helpful in increasing efficiency and saving time. All the participants gave a rank of 4 or more to this question, showing that the digital twin system would help in saving time of a cloud data centre administrator. The final question in this section is regarding the satisfaction of the digital twin system. Question Q8 was once again rated highly by all the participants, demonstrating that they were satisfied with the digital twin system.

\vspace{-0.5cm}
\begin{table}[h]
\scriptsize
\centering
\caption{\label{tab:results} Survey style questions answers on a Likert scale of 1-5.}
\begin{tabular}{ll|l|l|l|l|l|l|l}
\hline
\multicolumn{1}{|l|}{Participants} & \multicolumn{1}{l|}{Q1} & \multicolumn{1}{l|}{Q2} & \multicolumn{1}{l|}{Q3} & \multicolumn{1}{l|}{Q4} & \multicolumn{1}{l|}{Q5} & \multicolumn{1}{l|}{Q6} & \multicolumn{1}{l|}{Q7} & \multicolumn{1}{l|}{Q8} \\ \hline
\multicolumn{1}{|l|}{P1}           & \multicolumn{1}{l|}{4}  & \multicolumn{1}{l|}{1}  & \multicolumn{1}{l|}{4}  & \multicolumn{1}{l|}{5}  & \multicolumn{1}{l|}{4}  & \multicolumn{1}{l|}{5}  & \multicolumn{1}{l|}{5}  & \multicolumn{1}{l|}{4}  \\ \hline
\multicolumn{1}{|l|}{P2}           & \multicolumn{1}{l|}{4}  & \multicolumn{1}{l|}{3}  & \multicolumn{1}{l|}{4}  & \multicolumn{1}{l|}{4}  & \multicolumn{1}{l|}{3}  & \multicolumn{1}{l|}{4}  & \multicolumn{1}{l|}{5}  & \multicolumn{1}{l|}{4}  \\ \hline
\multicolumn{1}{|l|}{P3}           & \multicolumn{1}{l|}{5}  & \multicolumn{1}{l|}{2}  & \multicolumn{1}{l|}{3}  & \multicolumn{1}{l|}{2}  & \multicolumn{1}{l|}{4}  & \multicolumn{1}{l|}{4}  & \multicolumn{1}{l|}{4}  & \multicolumn{1}{l|}{5}  \\ \hline
\end{tabular}%
\vspace{-0.5cm}
\end{table}

The final section of the interview consisted of four questions as listed in the bottom section of Table~\ref{tab:Qs}. These questions were open-ended to give deeper insights and feedback into the digital twin system. The first question in this section is question Q9, the purpose of this question is to gauge if the participants would like any change to be made to the digital twin system or/and add any additional features. Each of the participants had different responses to this question, starting with P1, the participant recommended showing more data about the VMs, hovering over the boxes to show data like name of the instance, and a search bar to search and highlight a specific VM (box) on the system. P2 also answered that they would like to see more data represented in the system; data such as which user the VM (box) belongs. Also, they said that the current plates were homogeneous, a heterogeneous plate configuration would be better as it would show more data about the hypervisors. P3 answered this question by stating they would like to see a VM's history in the digital twin system. Having this history available would help the cloud data centre administrator make better decisions about VMs' placement on different hypervisors during different periods. For example, in a community cloud data centre in a university, a hypervisor might be overloaded every year during the exam period but during the semester break its load is below average, having this historical data can help the administrator make better decisions to make the system more efficient and save power.

The next question Q10 was regarding the negative aspects of the digital twin system. P1 said that the system needs to show error messages in case something goes wrong. P2 noted that the digital twin maintenance could be problematic, and the digital twin system cannot replicate all the data available on the OpenStack dashboard. P3 stated that they would like to see more metrics before moving a VM from one plate to another. Knowing the metrics would help avoid the hypervisors getting crowded and increase efficiency and server utilisation. 

The following question Q11 was to find out all the positive aspects of the digital twin. P1 stated that the digital twin was easy to use, faster than searching for information on the dashboard as it is visualised. It is easy to find an empty hypervisor for migration, and all in all, it provides a good user experience. P2 noted that the digital twin simplifies the cloud data centre's monitoring and management, easy for a user to use the whole system. It helps with the simplification of cloud data centre monitoring and management. Lastly, P3 answered that digital twin makes migration and management easy with lots of potential in the future. The migration aspect would help spread the load to different hypervisors by viewing which hypervisor has less load on them or by checking the energy consumption based on colour density.

The final question of the interview was regarding any additional feedback. Feedback given by P1 was an extension to the project to add the digital twin as a plug-in into OpenStack. P2 stated to show more colours and add more detail to the hypervisor plate, they said instead of keeping a single colour for the plates change them to red or green depends on the state up or down. Additionally, P2 stated to add pop-up windows to display more detail regarding the projects. Lastly, P3 suggested adding data from a \textit{ceilometer} to add more data regarding humidity and temperature of the real-world server rack. Another recommendation of P3 was to integrate the digital twin with \textit{Ralph3}, which is an asset management system for data centres; it gives a different type of visualisation compared to the digital twin. 

\section{Discussion and Future Directions} 

The key takeaway of the qualitative study according to the participants' feedback indicates that the digital twin system is user friendly and easy to use. More importantly, the digital twin system helps with increasing efficiency and resource management decisions, such as looking at the plates and knowing which hypervisors have less load on them or overloaded. The digital twin can be extended and improved by adding more details and giving more details to the user, adding more functionality and integration with ceilometer and ralph3. Another key takeaway is that the digital twin cannot replace the OpenStack dashboard as a monitoring tool, not its intended purpose. Openstack dashboard offers a great deal of information that cannot be all incorporated in a digital twin Because of the limitation of the 3D model. However, digital twin provides high-level views of the data centre resources and its status based on OpenStack information which is hard to grasp using the dashboard. It can be used as an extension to the OpenStack dashboard and makes it more user friendly, easier to use, and more efficient monitoring and management tool for data centre admins.

The threats for validation of this study are that there were only three participants involved. For evaluations of the future study, we will add more participants and have a more extensive study. The results are currently promising and suggest that the Data Centre Digital Twin maybe a useful tool for more efficient monitoring and management of complex cloud data centres.

The feedback given by the participants of the study helped fuel the future outlook of the project. The following are the directions for future works of the research. Firstly, adding a VR (virtual reality) component to the digital twin system. VR would give a different visual sense to the digital twin, using a VR headset to view in detail the digital twin in 3D and move around it to check the different components. One functionality that can be added is when the user clicks on a plate, all the boxes on the plate animate and moves to the space around the server rack where they are spread out in order. Doing this would make viewing the digital twin components easier. Other functionalities are showing error messages, having a search bar to make it easier to search VMs, hovering a mouse over a box would display more details. Also displaying the history of VMs and hypervisors in the model, making the digital twin a plugin on the OpenStack and using a ceilometer to get the real-world server rack's temperature and humidity, for instance, the temperature can be shown as a red to a green hue. Another additional feature would be to show the network traffic and how much traffic is directed to each server. Lastly, this singular digital twin can be multiplied into multiple server racks beside each other, common in data centres.

\section{Conclusions} This paper presented the instance of a digital twin built for OpenStack-managed cluster of servers. The system was designed to help cloud data administrators better monitor and manage different cloud data centre aspects. A 3D model was created to visualise different aspects such as VMs which are boxes in the system and hypervisors, which are plates. The digital twin also allows the user to use a couple of interactive features. To evaluate the effectiveness of the digital twin and gather feedback, a qualitative study was conducted where three participants were chosen who were experts in the field of cloud data centres and OpenStack. The participants were asked a series of questions to evaluate how the digital twin system performs compared to the OpenStack dashboard. The feedback was concise and broad, giving new insights into ways that the digital twin outperforms the dashboard to be more user-friendly and less time consuming to navigate.


%
%
%
%

\bibliographystyle{splncs04}
\bibliography{samplepaper}

\end{document}